\documentstyle[twocolumn,short,epsfig]{jpsj}

\title{Degeneracy of Ground State in 
Two-dimensional Electron-Lattice System}

\author{
Tetsuya {\sc Hamano} and Yoshiyuki {\sc Ono}
}

\inst{
Department of Physics, 
Toho University, Miyama 2-2-1, 
Funabashi, Chiba 274-8510
}

\recdate{January 31, 2001}
\kword{Peierls distortion, SSH-type electron-lattice 
interaction, degeneracy of ground state, square lattice}

\newcommand{\vv}[1]{\mbox{\boldmath{$#1$}}}

\begin{document}
\sloppy
\maketitle

In a recent paper,~\cite{rf:1} we have reported the ground state of 
a two dimensional system described by a SSH(Su-Schrieffer and Heeger)-type 
Hamiltonian with a half-filled electronic band. The 
Hamiltonian~\cite{rf:2} is given by 
\begin{eqnarray}
H = &-& \sum_{i,j,\sigma}[(1 - x_{i,j})(C_{i,j,\sigma}^{\dagger} 
C_{i+1,j,\sigma} + {\rm h.c.}) \nonumber \\
&+& (1 - y_{i,j})(C_{i,j,\sigma}^{\dagger} 
C_{i,j+1,\sigma} + {\rm h.c.})] \nonumber \\
&+& \frac{1}{2 \lambda} \sum_{i,j}[x_{i,j}^2 + y_{i,j}^2], \label{eq1}
\end{eqnarray}
where the field operators $C_{i,j,\sigma}$ and $C_{i,j,\sigma}^\dagger$ 
annihilate and create an electron with spin $\sigma$ at the site $(i,j)$ on 
a square lattice subject to the periodic boundary conditions (PBC)
, and the energy 
is scaled by the transfer integral of the regular lattice, the bond variables 
$x_{i,j}$'s and $y_{i,j}$'s in $x$- and $y$-directions, respectively, being 
scaled appropriately to involve the electron-lattice coupling coefficient, 
$\lambda$ representing the dimensionless coupling constant. In present work 
we fix the coupling constant at $\lambda=0.3$, because we are interested only 
in the properties of the ground state which would not depend on the value of 
$\lambda$. The ground state is accompanied by lattice distortions which have 
Fourier components not only with the nesting vector $\vv{Q}=(\pi/a , \pi/a)$ 
but also with many other wave vectors parallel to $\vv{Q}$,~\cite{rf:1} 
\begin{eqnarray}
x_{i,j} & = & x_{i,j}^{\rm N} + x_{i,j}^{\rm O} \label{eq2}, \\
x^{\rm N}_{i,j} &=& X_{\frac{\pi}{a},\frac{\pi}{a}} (-1)^{i+j} \label{eq3}, \\ 
x^{\rm O}_{i,j} &=& \sum_{0< q < \frac{\pi}{a}} 
( X_{q,q} {\rm e}^{{\rm i} q (i+j) a} 
+ X^{\ast}_{q,q} {\rm e}^{- {\rm i} q (i+j) a}), \label{eq4} 
\end{eqnarray}
and a similar expression for $y_{i,j}$. Here $a$ represents the lattice 
constant. 
Furthermore, the pattern of these distortions is found not unique in 
the lowest energy state, although the structure of the electronic energy 
spectrum and the total energy of the system are the same.~\cite{rf:1}
In this note we discuss the degrees of degeneracy of the lowest energy state 
for the model described by eq. (\ref{eq1}). 

\begin{figure}[htb]
\hspace*{15mm}
\epsfig{file=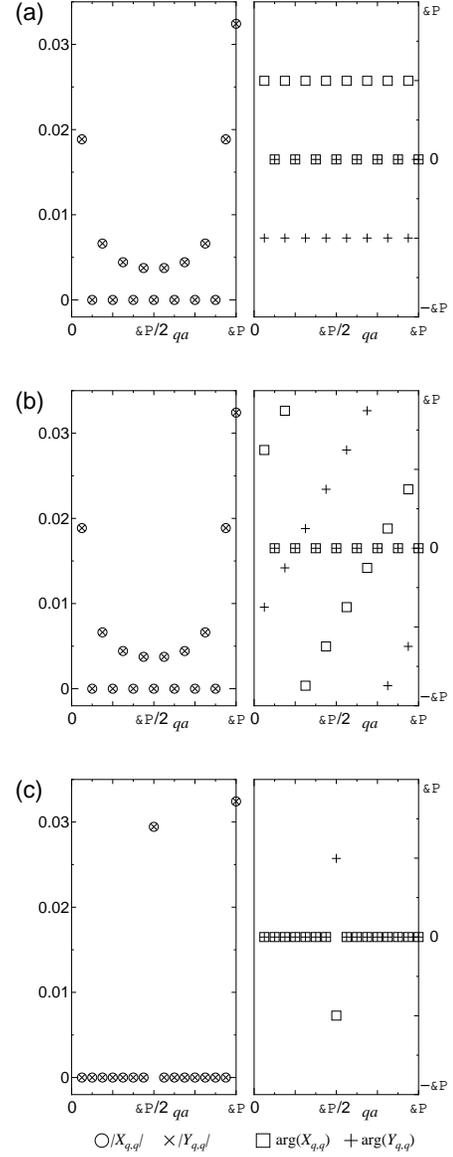,height=15.6cm} 
\caption{
The wave number dependence of the Fourier components of 
the lattice distortion. The system size is $32 \times 32$. 
For the definition of $X_{q,q}$ and $Y_{q,q}$, see eqs.~(\ref{eq2}) to 
(\ref{eq4}).} \label{fig1}
\end{figure}
It is clear that there is trivial degeneracies due to the symmetry of the 
system. Three examples of the distortion patterns in the ground state is 
shown in terms of the wave number dependence of the Fourier components. 
In each graph, the amplitudes of $X_{q,q}$ and $Y_{q,q}$ are plotted as 
functions of $q$ ($0 < q \le \pi/a$) on the left-hand side, while their 
arguments are shown on the right-hand side. 
The amplitudes are exactly the same in (a) and (b) of 
Fig.~\ref{fig1} though the arguments look quite different. It is not 
difficult to understand that the difference between Figs.~\ref{fig1}(a) and 
\ref{fig1}(b) can be explained by the translation symmetry of the system. 
The behavior of the amplitudes in Fig.~\ref{fig1} (c) is completely different 
from that in Figs.~\ref{fig1}(a) and \ref{fig1}(b). This fact indicates that 
there exist non-trivial degeneracy in the ground state. From Fig.~\ref{fig1} 
along with many other data which are not shown here, we can summarize 
the characteristics 
of the distortion patterns in the ground state as follows; (1) the amplitude 
of $X_{q,q}$ is always equal to that of $Y_{q,q}$, (2) the argument of 
$X_{q,q}$ is equal to that of $Y_{q,q}$ or differs from that of $Y_{q,q}$ 
by $\pi$, and (3) the values of $X_{\frac{\pi}{a},\frac{\pi}{a}}$ and 
$Y_{\frac{\pi}{a},\frac{\pi}{a}}$ are common for all the patterns. 
Thus the variety of the distortion patterns in the ground state can be said 
to be determined by the values of $X_{q,q}$'s with $0<q<\pi/a$. 

In order to discuss the degrees of degeneracy of the ground state, we 
assume first that the Fourier components of the distortion are restricted to 
the diagonal ones with $q_x=q_y$ as numerically confirmed in the previous 
paper,~\cite{rf:1} and that the displacements $y_{i,j}$ in $y$-direction 
are automatically determined if the displacements $x_{i,j}$ in $x$-direction 
are fixed. Then the number of variable describing the distortion pattern can 
be reduced from $2\times (N-1)^2$ to $N-1$ for a system with the size 
$N\times N$. 
It will be allowed to choose $x_{i,J}$ ($i=1,2,\cdots,N-1$) with a fixed 
$J$ as a set of $N-1$ independent variables. 
(Note the constriction $\sum_i x_{i,J} = 0$ due to the PBC.)

\begin{figure}[htb]
\hspace*{15mm}
\epsfig{file=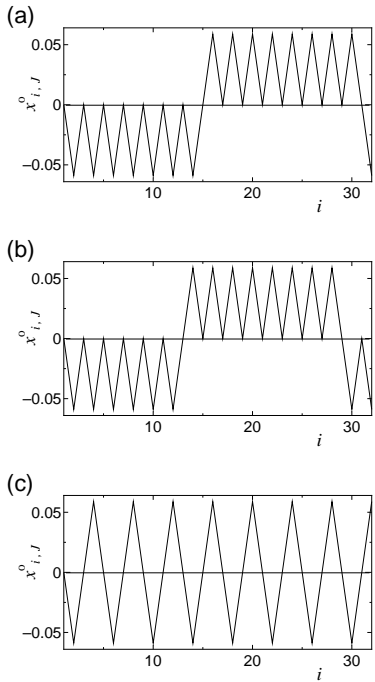,height=9.2cm} \vspace*{1mm} 
\caption{$x^{\rm O}_{i,J}$ for a fixed $J$ is plotted as a function of 
$i$. (a), (b) and (c) correspond to those of Fig.~\ref{fig1}. The value of 
$J$ is taken to be 1 in these cases. Changing $J$ leads to a translation 
of the pattern in the direction of $x$ (or $i$).} \label{fig2}
\end{figure}

Figures \ref{fig2}(a) to \ref{fig2}(c) show the distortion $x^{\rm O}_{i,J}$ 
for a fixed $J (=1)$ as a function of $i$, which is constructed from 
$X_{q,q}$'s with $0<q<\pi/a$ of the corresponding figures of Fig.~\ref{fig1} 
as in eq.~(\ref{eq4}).~\cite{rf:3} It will be clear that Figs.~\ref{fig2}(a) 
and \ref{fig2}(b) can be completely overlapped by a simple translation along 
$i$-axis.  On the other hand Fig.~\ref{fig2}(c) is quite different from them. 
Nevertheless there exists a common feature among the three patterns. Namely, 
(I) $x^{\rm O}_{i,J}$ vanishes at every second site, and (II) the number of 
sites at which it is positive is the same as the number of sites with 
negative value. Furthermore (III) the absolute values of the positive and 
negative $x^{\rm O}_{i,J}$'s are equal to each other and common to all the 
three patterns. It will be plausible to assume that the absolute value is 
determined by the dimensionless coupling constant $\lambda$. All the numerical 
data we have obtained show that there is no rule in the order of positive and 
negative $x^{\rm O}_{i,J}$'s. Thus we can conjecture that the degeneracy of 
the ground state is given by different order of positive and negative 
$x^{\rm O}_{i,J}$'s. If $N$ is a multiple of four, then $x^{\rm O}_{i,J}$ will 
be zero at a half of $N$ sites ($i=1, \cdots, N$) and will be positive at a 
half of remaining $N/2$ sites and negative at remaining $N/4$ sites. 
The degree of non-trivial degeneracy 
except the trivial degeneracy due to arrangement of zero points 
might be said to be $_{N/2}{\rm C}_{N/4}$ 
($\simeq N^{N/4}$ for $N \gg 1$).

In order to check whether the above-mentioned rule, (I) to (III), for the 
distortion pattern is a sufficient condition for the ground state, we 
calculate, for the system size $32\times 32$, the total energy for all the 
lattice distortion patterns satisfying the above-mentioned rule, whose number 
is $_{16}{\rm C}_8 = 12,870$. 
For the amplitude of nonvanishing $x^{\rm O}_{i,J}$, we use the value shown 
in Fig.~\ref{fig2}. In all the cases the resulting energy is found 
equal to the minimum energy. Thus the above-mentioned rule can be regarded as 
a sufficient condition. 

It is not easy to show the above rule is a necessary condition for the ground 
state distortion pattern. However we can confirm it numerically to some extent. 
As discussed in the previous paper,~\cite{rf:1} we may believe that the ground 
state distortion for the present model has only the Fourier components with 
$q_x=q_y$. In this situation we can reduce the 2D Schr\"odinger equation to a 
1D equation in the momentum space.~\cite{rf:1} In order to perform the 
calculations finding the ground state effectively, we adopt this formulation. 
We start the iterative 
calculation by giving random numbers for the diagonal Fourier components of 
the distortions $x_{i,j}$'s and $y_{i,j}$'s and continue the iteration up to 
the 100-th step. 
The number of runs carried out is 100,000. Out of all the runs we 
could obtain 20,445 minimum energy states,~\cite{rf:4} some of them are 
equivalent within the translational shift. We could confirm that the above 
rule is satisfied by all the lowest energy states obtained in these 
calculations. Since we find no counter example against the above rule in a 
number of calculations starting from random initial inputs, it may be allowed 
to say that the above-mentioned rule is a necessary condition for the ground 
state distortion pattern of the present model. 
We have confirmed the above rules for different system sizes, 
$N=8,12,16,32,64$ and 128, though the sample number for $N=64$ and 128 are 
rather restricted, and also for tilted square systems~\cite{rf:5} 
satisfying PBC's. 

The degeneracy treated in this paper is considered to be due to the highly 
symmetric structure of the model. The Peierls distortions in two-dimensional 
anisotropic systems are now under investigation. The results will be published 
elsewhere.



\begin{thebibliography}{99}
\bibitem{rf:3} If we increase $J$ by one, the pattern shifts to larger 
$i$ direction by one.
\bibitem{rf:4} Since there are other local minima, it is not possible to 
get the lowest energy state in all of the runs. A typical local minimum is 
the one studied by Tang and Hirsch (Phys. Rev. B{\bf 37} (1988) 9546), which 
has only the Fourier component with the nesting wave vector.
\end{thebibliography}
\end{document}